\documentclass{PoS}
\usepackage{amsmath}
\usepackage{slashed}
\usepackage{amssymb}
\usepackage{bm}
\usepackage{caption}
\usepackage{latexsym}
\usepackage{graphicx}
\usepackage{psfrag,fancyhdr,epsfig}

\title{Aspects of the Flavour Expansion Theorem}

\ShortTitle{Aspects of the Flavour Expansion Theorem}

\author{\speaker{M. Paraskevas}%
         \\
        University of Ioannina\\
        E-mail: \email{mparask@grads.uoi.gr}}

\abstract{The \emph{Flavour Expansion Theorem}, which has been recently proposed as a more general and elegant algebraic method, for the derivation of the commonly used \emph{Mass Insertion Approximation}, is revisited. The theorem is reviewed, with respect to its straightforward applications in Flavour physics, and compared against the standard diagrammatic flavour basis techniques, in cases where the latter become inadequate. }

\FullConference{18th International Conference From the Planck Scale to the Electroweak Scale \\
                 25-29 May 2015\\
                 Ioannina, Greece }

\begin{document}

\section{Introduction}
In perturbative Quantum Field Theories (QFT), the Lagrangian plays a fundamental role, since it contains all information regarding the dynamics of a given system. In particle physics, the Lagrangian is typically invariant under a set of symmetries whose eigenstates correspond to the physical states of the theory. Expressing the Lagrangian in the physical basis
 has many advantages, including the apparent identification of the states of the theory with physical particles. Nevertheless, one is always allowed to perform  unitary transformations, which respect the symmetries of the theory, and express the Lagrangian in any other basis, resulting to a different but equivalent description of QFT.

 At first sight, expressing the Lagrangian in another basis, besides the physical one, may seem an unnatural choice with no physical purpose. However, there are cases where an underlying mechanism for certain phenomena becomes clear only in an ``unphysical" basis (\emph{e.g., suppression of transition amplitudes, cancellation between diagrams, etc.}). This is typically the scenario for symmetry effects, in theories with a symmetry breaking mechanism. The QFT description in terms of the physical eigenstates (\emph{i.e. mass basis QFT}) is usually inadequate to explain phenomena related to the initial symmetry currents. For that purpose, the description in terms of the initial symmetry eigenstates (here collectively referred to as \emph{flavour basis QFT}) can be considerably more illuminating. 
 
The standard method for flavour basis QFT, is a well known diagrammatic technique called the \emph{Mass Insertion Approximation} (MIA) \cite{Hall:1985dx,Gabbiani:1996hi,Misiak:1997ei}. Its derivation follows the footsteps of mass basis diagrammatic QFT,(\emph{i.e., propagators, vertices)} but with the significant difference that it is applied on the theory before mass diagonalization and field redefinitions. In  case of theories with an embedded symmetry breaking mechanism, this unphysical basis corresponds to the basis of the initial symmetry, what here referred to, as flavour basis. In MIA, all interactions are treated in flavour basis, as they would be treated in mass basis. One reads cubic and quartic interactions directly from the Lagrangian and extracts the corresponding 3,4-point vertex rules, with the couplings obviously defined in flavour basis. Every interaction coupling in MIA approach is related to its corresponding mass basis coupling through the unitary transformations, associating the two eigenstate bases.   

\begin{figure}[t]
{\center{
    \includegraphics[trim=-3.cm 3.cm 0.0cm 2.8cm, clip=true, totalheight=0.12\textwidth]{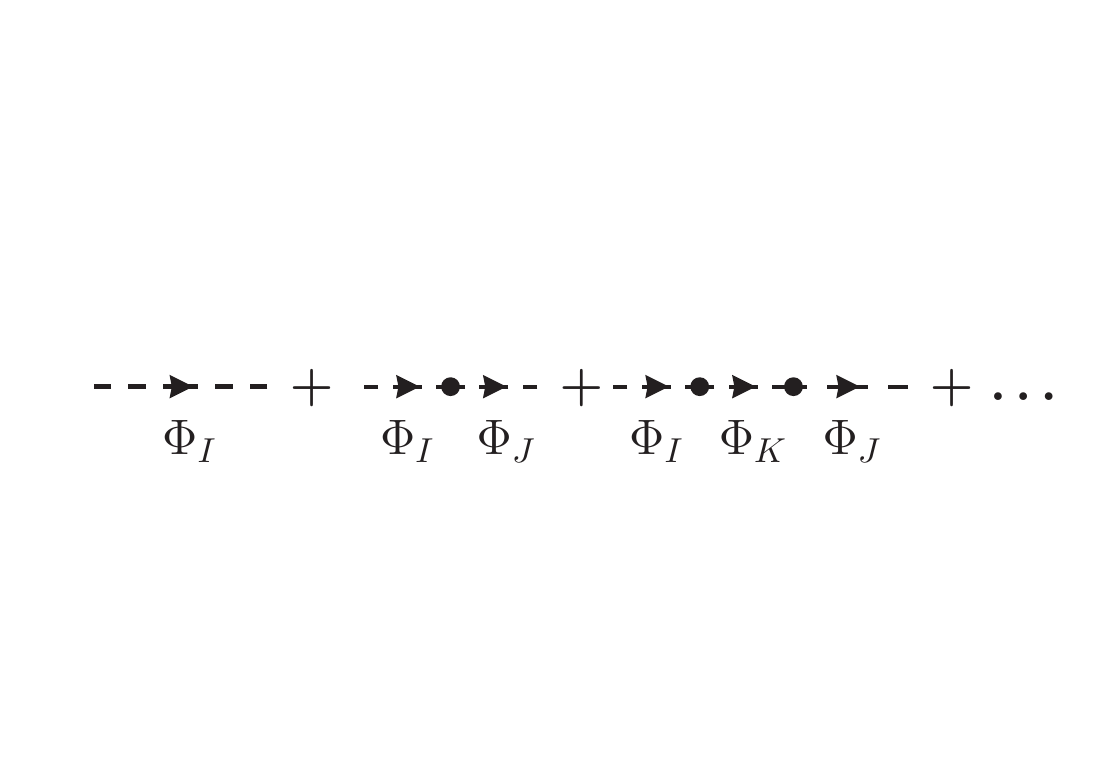}}}\vspace{0.3cm}
\caption{\emph{The Flavour propagator in the MIA. The infinite sum of propagators - mass insertions, converges to a Hermitian matrix function $\Delta_p(\mathbf{M^2})$, which is associated to the mass basis propagator, through the corresponding unitary transformation, i.e., $\Delta_p(\mathbf{m^2}) = \mathbf{U^\dag}\,\Delta_p(\mathbf{M^2})\,\mathbf{U}$.} }\label{fig-flavourprop}    
\end{figure}

 Although mathematically consistent, expressing the Lagrangian in terms of unphysical degrees of freedom has a price to pay. This is reflected in the existence of quadratic mixing interactions (2-point vertices, commonly referred to as \emph{mass insertions}) which, conversely, are always absent in the mass basis description. As a result, even in exactly solvable theories (\emph{i.e., up to quadratic interactions}), the propagator in flavour basis becomes an iterative expansion in terms of mass insertions, as shown in fig.\ref{fig-flavourprop}. The situation becomes considerably more cumbersome when  higher order interactions are present, as well. For each diagram in mass basis, there is an infinite number of diagrams in flavour basis, due to the iterative expansion of each flavour propagator in the diagram. Nevertheless, as long as small mass insertions are considered, the expansion is expected to converge fast and thus, by keeping only the first few terms, one can obtain an approximate but useful result for the qualitative interpretation of certain effects. The usefulness of this method, relies to the fact that mass insertions, which reflect the flavour violation of the theory, are treated as couplings in MIA. As with every coupling in a perturbative QFT, they can only appear as bare factors in a diagram calculation, thus associating the contribution of a given diagram with the flavour violation of the theory, in a straightforward manner.\footnote{Alternatively, it is  possible to obtain the exact flavour basis result, directly from the mass eigenstate calculation, by expressing all mass basis parameters (masses, unitary matrices, couplings) in terms of flavour basis parameters, exclusively. The result is consistent, but practically useless since  flavour violation is not factored out, but instead deeply encoded within the arguments of the respective loop-function.}

 As one can easily imagine, in theories with complicated flavour structure, the diagrammatic MIA method becomes a complex and error-prone process. On the other hand, the recently developed Flavour Expansion Theorem (FET) \cite{FET,Rosiek:2015jua}, offers an alternative derivation of the MIA result, without the difficulties that unavoidably arise from the use of diagrammatic techniques, in flavour basis QFT calculations. In addition, embedding the MIA within the solid mathematical framework of matrix analysis \cite{bhatia97,0521386322}, provides with various new benefits. As a purely algebraic method, FET allows, for a discussion on theoretical aspects of flavour basis QFT (\emph{i.e. convergence}), which  were inaccessible by any diagrammatic approach, and also for the consistent derivation of the MIA result even in cases where there is no clear diagrammatic picture.

\section{The Flavour Expansion Theorem: A brief review} 
 The theorem, applied first in \cite{Dedes:2014asa}, but formulated and proved in \cite{FET}, suggests that any analytic function of a Hermitian matrix $f(\mathbf{A})$ can be expanded polynomially in terms of the off-diagonal elements of the matrix, a property which will be shown to have  an intimate connection with the MIA expansion, as discussed in what follows. In more detail, if $f(\mathbf{A})$ is an analytic Hermitian matrix function, then any element of this matrix function will be given by the expansion,
\begin{align}
f({\bf A})_{IJ} & =\ \delta_{IJ} f(A_I)\ +
\ \hat{A}_{IJ}\:f^{[1]}(A_I,A_J)\ +\ \sum_{K_1}
\hat{A}_{I{K_1}}\hat{A}_{{K_1}J}\:f^{[2]}(A_I,A_{K_1},A_J)
\nonumber\\
&\ + \sum_{K_1,K_2}\hat{A}_{I{K_1}}\hat{A}_{{K_1}{K_2}}\hat{A}_{{K_2}J}\:f^{[3]}(A_I,A_{K_1},A_{K_2},A_J) \label{fet}
\ +\ \dots \ , 
\end{align}
following the matrix decomposition into diagonal and (purely) off-diagonal part, of the form,
\begin{align}
\mathbf{A} &= \mathbf{A_0} +\mathbf{\hat{A}}\nonumber\\
A_{IJ}&= A_I \delta_{IJ} +\hat{A}_{IJ}\;,\quad \hat{A}_{II}\equiv 0 \label{Adecom}
\end{align}
and the mathematical formalism of the \emph{Divided Differences (DD)} \cite{de2005divided}  , obtained through the \emph{symmetric}, recursive division process,
\begin{align*}
f^{[0]}(x_0)&\equiv f(x_0)\,,\:\\
f^{[1]}(x_0,x_1)&\equiv \frac{f(x_0)-f(x_1)}{x_0-x_1}\,,\:\\
f^{[2]}(x_0,x_1,x_2)& \equiv\frac{f^{[1]}(x_0,x_1)-f^{[1]}(x_0,x_2)}{x_1-x_2}\,,\:\\
f^{[k+1]}(x_0,\dots ,x_{k+1})&\equiv\frac{f^{[k]}(x_0,\dots,x_{k-1},x_k)-f^{[k]}(x_0,\dots,x_{k-1},x_{k+1})}{x_k-x_{k+1}}\,,\\
\lim_{\{ x_0,\dots ,x_m\} \to \{ \xi,\dots,\xi \}}
f^{[k]}(x_0,\dots,x_k)&=\frac{1}{m!}\frac{\partial^{m} }{\partial
  \xi^{m}}f^{[k-m]}(\xi,x_{m+1}\dots,x_k)\quad .
\end{align*}
The last equation, corollary of the \emph{mean value theorem for DD}, suggests a uniform treatment for the matrix expansion, in case of degenerate or identical arguments.
  
It is not hard to imagine the straightforward applications of the above theorem in flavour physics. With FET one can translate a mass basis expression for a transition amplitude into an expansion in terms of mass insertions, instantly \cite{Rosiek:2015jua}. For example, a mass basis diagram (or subgraph) at any loop, involving scalars with non-trivial flavour structure,  can be expressed in the general form,  \footnote{Here, all other arguments, irrelevant to flavour (\emph{e.g., external momenta, other masses}) have been suppressed.}
\begin{align}
\sum_i U_{Ii} f(m_i^2) U^{\dag}_{iJ} &= f(\mathbf{M^2})_{IJ}\nonumber\\ &= \delta_{IJ} f(M_I^2)\ +
\ \hat{M}^2_{IJ}\:f^{[1]}(M_I^2,M_J^2)\ +\dots \ ,\end{align}
where the notation, following the assumption of eqn.(\ref{Adecom}), implies a matrix decomposition of the form,
\begin{align}
(M^2)_{IJ} &= (M^2)_{II} \delta_{IJ} + (M^2)_{IJ}\Big|_{I\neq J}\equiv M^2_I \delta_{IJ} + \hat{M}^2_{IJ}
\end{align}
The elements of the Hermitian Flavour (squared) Mass Matrix ($\mathbf{M^2}\equiv \mathbf{A}$), are associated to the mass eigenvalues, $m_i^2$, through the unitary rotation, 
$$\mathbf{U \:m^2\, U^\dag} = \mathbf{M^2}\ . $$
As has been shown in \cite{FET} analogous, FET-treatable expressions are obtained from diagrams involving fermions (or gauge bosons) as well.

The generalization to the multivariable case is always straightforward. For example, the FET expansion for two variables will have the form,
\begin{align}
\sum_{i,j} U_{Ii} V_{Kj}\; f(m_i^2,m_j^{\prime \,2})\; V^\dag_{jL}U^{\dag}_{iJ} &= f(\mathbf{M^2},\mathbf{M^{\prime \,2}})_{(IJ),(KL)} \nonumber\\&= \delta_{IJ} f(M_I^2,\mathbf{M^{\prime \,2}})_{KL}\ +
\ \hat{M}^2_{IJ}\:f^{[1]}(M_I^2,M_J^2,\mathbf{M^{\prime \,2}})_{KL}\ +\dots \nonumber\\
 &= \delta_{IJ}\delta_{KL} f(M_I^2,M_K^2)\ \nonumber \\
 &\ +
\ \hat{M}^2_{IJ}\delta_{KL}\:f^{[1]}(M_I^2,M_J^2,M_K^2)\ +
\ \delta_{IJ}\hat{M}^2_{KL}\:f^{[1]}(M_I^2,M_K^2,M_L^2)\nonumber \\& \ +\ \dots \quad  .\end{align}
As in the single variable expansion, summations over internal indices will only arise for terms at second order DD \emph{(i.e., $\sim f^{[2]}$)}, and higher.

\section{FET vs diagrammatic MIA}
Some comments on theoretical aspects of FET are in order:
  
\begin{itemize}
\item 
\emph{The general FET formula in eqn.(\ref{fet}), unifies all possible Mass Insertion Approximation treatments of QFT, in a single algebraic expression.}
\end{itemize}

The decomposition used in eqn.(\ref{Adecom}), follows the implicit assumption that $\mathbf{\hat{A}}$ is purely off-diagonal. As a result, terms proportional to $\hat{A}_{II}$ will necessarily vanish in the summations of eqn.(\ref{fet}). However, such an assumption is not a necessary ingredient for the formal proof of FET and therefore this general expansion formula will still hold even when $\mathbf{\hat{A}}$ carries non-vanishing diagonal elements as well. 

To explain this statement in more detail, FET is categorized into three major classes, corresponding to the (only) three general assumptions on the decomposition of the Hermitian Matrix $\mathbf{A}$. The connection to flavour physics is easily understood, by substituting  $\mathbf{A}\rightarrow \mathbf{M^2}$, in the following expressions. The three distinct cases for the decomposition in eqn.(\ref{Adecom}) are:
\begin{subequations}
\begin{align}
& \hspace{4cm}A_{IJ} = A_I\delta_{IJ} + \hat{A}_{IJ}\ ,\nonumber\\
\hat{A}_{II} &= 0\ , \qquad\qquad \emph{(Standard MIA for bosons - massive propagator description)}\ ,\label{massMIA}\\
\hat{A}_{II} & \neq 0\ ,\ A_I \neq 0\ , \qquad  \emph{(Mixed MIA - (infinite) mixed propagator descriptions)}\ ,\label{mixMIA}\\
A_I &= 0\ , \qquad \; \emph{(Standard MIA for fermions - massless propagator description)}\ ,\label{nomassMIA} 
\end{align}
\label{Adecomclass}
\end{subequations}

Any choice above, is legitimate, as long as all criteria for convergence and analyticity in FET expansion are fulfilled. Furthermore, every matrix decomposition assumption for FET, is in one to one correspondence with a certain MIA diagrammatic treatment. For example, the FET expansion obtained under the assumption of eqn.(\ref{massMIA}), will be shown to be, order by order equal to the expansion obtained from diagrammatic MIA, under the assumption that the diagonal part of quadratic mixing is fully absorbed in the definition of the propagator and the 2-point vertices are purely off-diagonal (\emph{i.e. $\hat{M}_{II}^2 = 0$}). Analogous correspondence between FET and diagrammatic MIA can be shown to exist in each case.

Different decomposition assumptions in eqs.(\ref{Adecomclass}), result into different explicit expansions, which are not equal order by order, since the arguments (\emph{i.e. $A_I,\hat{A}_{IJ}$}) in eqn.(\ref{fet}), are different in each case.  However, all expansions are equivalent, since when convergent, they always converge to the same Hermitian matrix function, $f(\mathbf{A})$, but with different convergence rates. In cases where all three descriptions are legitimate (\emph{i.e., analyticity/convergence issues are fulfilled}), the ``massive propagator description" in eqn.(\ref{massMIA}) will be the preferable method for physical calculations, due to its high rate of convergence, and due to the fact that the expansion parameter ($\hat{A}_{IJ}\equiv $ mass insertion) carries, the flavour violation effect of the theory.  

\begin{figure}[t]
\vspace{-1.5cm}{\center{
    \includegraphics[trim=0cm 2cm 0cm 0.cm, clip=true, totalheight=0.28\textwidth]{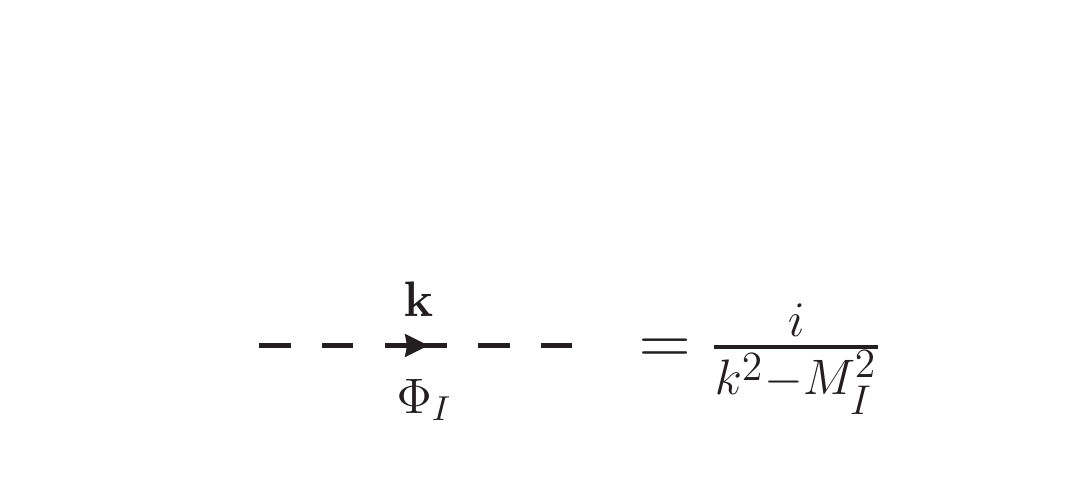}}
    \includegraphics[trim=2.8cm 2cm 0cm 0.cm, clip=true, totalheight=0.28\textwidth]{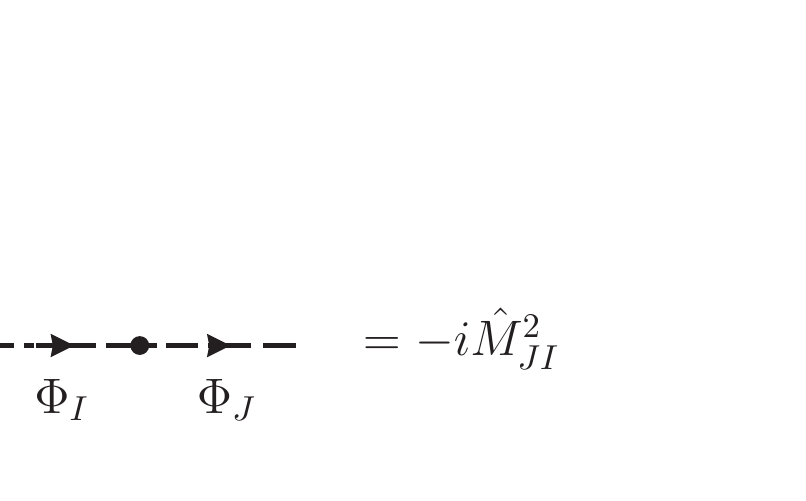}}
    \caption{\emph{Feynman rules for the MIA. In the massive MIA description, the 2-point vertices of the theory (i.e., mass insertions) are purely of diagonal ($\hat{M}^2_{II} = 0$). In interacting theories, for each n-point vertex rule in mass basis $(n\geq 3)$, there is a corresponding n-point vertex rule in MIA, with the same structure, but with the flavour couplings defined in terms of the flavour eigenstates, $\Phi_I$.}  } \label{fig-MIArules}
\end{figure}

\begin{itemize}
\item \emph{The standard diagrammatic MIA expansion for bosons is equivalent and order by order equal to FET of type (\ref{massMIA}).}
\end{itemize}
It is a straightforward procedure, to apply the standard diagrammatic MIA for bosons, in an exactly solvable scalar theory- the generalization to higher order interactions, is always straightforward. By considering a flavour basis Lagrangian of the form 
$$\mathcal{L} = \sum_I\partial^\mu \Phi^\dag_I \partial_\mu \Phi_I - \sum_{IJ}(M^2)_{IJ} \Phi^\dag_I \Phi_J $$
one can instantly read the MIA diagrammatic rules of the theory, as shown in fig.{\ref{fig-MIArules}}. In the standard MIA for bosons, used here, the mass insertions have no diagonal part ($\hat{M}^2_{II} =0$). The full flavour propagator of the theory which is an infinite expansion over propagators and mass insertions, displayed in fig.\ref{fig-flavourprop}, can be summed over and expressed as a Hermitian  matrix function, by applying trivial matrix algebra techniques\footnote{The matrix identities $\sum_{n=0}^\infty \mathbf{A^n} = (\mathbf{I}-\mathbf{A})^{-1}$ and $(\mathbf{AB})^{-1}= \mathbf{B}^{-1}\mathbf{A}^{-1}$ are required. Convergence of the expansion is assumed to be satisfied.}. It will have  the form,    
$${i\over p^2 - M^2_{I}} \delta_{IJ}\ +\ {i\over p^2 - M^2_I} \Big(\hat{M}^2_{IJ} {1\over p^2 - M^2_J}\Big)+ \ \dots \  = \left({i\over p^2 - \mathbf{M^2}}\right)_{IJ} \equiv\Delta_p (\mathbf{M^2}) $$
The same expansion can be obtained with FET, under the decomposition assumption (\ref{massMIA}), but from a totally different starting point, namely the mass eigenstates propagator, as
\begin{align*}
&\sum_i U_{Ii} {i\over p^2-m_i^2} U^\dag_{iJ} =  \mathbf{U}{i\over p^2-\mathbf{m}^2} \mathbf{U^\dag} =  \left({i\over p^2 - \mathbf{M^2}}\right)_{IJ}\\ &\overset{FET}{=} {i\over p^2 - M^2_{I}} \delta_{IJ}\ +\ {i\over p^2 - M^2_I} \Big(\hat{M}^2_{IJ} {1\over p^2 - M^2_J}\Big)+ \ \dots \  
\end{align*}
Both expressions are matrix expansions in terms of the same parameter ($\mathbf{\hat{M}}^2$) and converge to the same Hermitian matrix function ($\Delta_p (\mathbf{M^2})$), thus they are equal order by order. It is also trivial to check that for any other consistent choice on the diagrammatic MIA feynman rules, the resulting expansion would also converge to the same Hermitian matrix function $\Delta_p (\mathbf{M^2})$.

Since the MIA expansion is essentially an iterative expansion over the flavour propagators of the theory, the presence of higher order interactions cannot affect this conclusion. In this more general case the Hermitian matrix function for the transition amplitude is $\Delta_p (\mathbf{M^2})\rightarrow f(\mathbf{M^2})$ where $f$ is some n-point correlation function (\emph{i.e., the full transition amplitude, a diagram or even a subgraph}).  Therefore, the conclusion is even more general. All consistent diagrammatic MIA expansions for any transition amplitude converge to the same Hermitian matrix function and can be obtained purely algebraically through the mass eigenstate result and a suitable FET treatment, following the corresponding matrix decomposition assumption.  

\begin{itemize}
\item \emph{The FET expansion can give a consistent MIA result even in cases where there is no clear diagrammatic picture - the massive MIA for fermions.}
\end{itemize}

In chiral theories, the standard diagrammatic MIA method follows the massless propagator description of eqn.(\ref{nomassMIA}). In more detail, in this approach, all quadratic interactions of the theory are treated as perturbative interactions, leading into a description where mass insertions (2-point vertices) have both diagonal and non-diagonal contributions, but the propagator is defined massless. In fact, although this approach is less attractive for practical calculations, since both the convergence rate is smaller and the flavour violation effect is not transparent (as compared to the massive MIA), it is the approach with the simplest and consistent diagrammatic picture. The reason for that is fundamental, since only massless propagators preserve chirality. 

Technically, the difficulty to define a massive fermion propagator in diagrammatic MIA, arises from the fact that the diagonal part of the fermion mass matrix, $\mathbf{M_0} P_L + \mathbf{M^\dag_0} P_R$, cannot be absorbed in the definition of massive propagators, since in general $\mathbf{M_0},\mathbf{M^\dag_0}$ are different. 
 Nevertheless, a consistent massive MIA treatment is still possible if one considers, equivalently, the flavour propagator as the direct sum of its four chiral projections \cite{FET}. The propagator is then expressed as the matrix function, 
\begin{align}
\Delta_p(\mathbf{M},\mathbf{M^\dag}) &=  \Delta_{LL}+\Delta_{RL}+\Delta_{RR}+\Delta_{LR}\nonumber \\
&= (\slashed{p}+\mathbf{M^\dag}) {i \over p^2 - \mathbf{MM^\dag}}\ P_L \ +\ (\slashed{p}+\mathbf{M}) {i \over p^2 - \mathbf{M^\dag M}}\ P_R\,.
\end{align}
Since both fractions, above, are essentially Hermitian Matrix functions of $(\mathbf{MM^\dag})$ or $(\mathbf{M^\dag M})$, it is trivial to FET-expand them under the matrix decomposition of eqn.(\ref{massMIA}) and obtain their massive MIA expansion. However, the corresponding diagrammatic MIA rules for such an expansion (\emph{i.e. propagators, 2-point vertices}) are far from obvious, and the diagrammatic picture is very subtle, even without considering higher order interactions.  

As already implied, the inconsistency between massive diagrammatic MIA and chiral theories is easily and elegantly resolved, by FET. The reason is that FET is a purely algebraic method, which requires no diagrammatic picture, in order to reproduce the MIA result. In fact its overwhelming power becomes clear when one realizes that it can act with the same efficiency both, on fundamental QFT objects such as propagators, or on complicated ones, such as n-point transition amplitudes at any loop order. As has been shown in \cite{FET}, any amplitude involving fermions with non trivial flavour structure will be described by one of the following general, FET-treatable forms,
\begin{align*}
U_{Ii}f(m^2_i)U^\dag_{iJ} &= f(\mathbf{M^\dag M})_{IJ}\\
V_{Ii}f(m^2_i)V^\dag_{iJ} &= f(\mathbf{M M^\dag})_{IJ}\\
U_{Ii} m_i f(m^2_i) V^\dag_{iJ} &= \sum_K M^\dag_{IK}f(\mathbf{M M^\dag })_{KJ}  = \sum_K f(\mathbf{ M^\dag M })_{IK} M^\dag_{KJ}\\
V_{Ii} m_i f(m^2_i) U^\dag_{iJ} &= \sum_K M_{IK}f(\mathbf{M^\dag M})_{KJ}  = \sum_K f(\mathbf{M M^\dag })_{IK} M_{KJ}
\end{align*}
each corresponding to one of the four chiral projections that any fermionic amplitude can be decomposed into. 

The physical consistency of this approach, is also verified by the fact that the FET expansion in this case, is an expansion in terms of the off-diagonal elements of the physically meaningful  Hermitian (squared) mass matrices $(MM^\dag)_{IJ} ,(M^\dag M)_{IJ}$. Although $\mathbf{M}$ or $\mathbf{M^\dag}$ are in general arbitrary complex matrices, their product is by default a Hermitian and semi-positive definite matrix, whose diagonal part is invariant under phase redefinitions of fields. Both Hermitian mass matrices share the same non-negative eigenvalues which correspond to the physical fermion (squared) masses of the theory and by being semi-positive definite, they also satisfy certain, crucial conditions for the convergence of the expansion \cite{FET}.\footnote{For a semi-positive definite Hermitian matrix $\mathbf{A}$, the properties $A_I\geq 0,\, \sqrt{A_I A_J}\geq |\hat{A}_{IJ}|$, always hold for any $I,J$.}

\section*{Acknowledgements}
The author wishes to thank  A.Dedes, J.Rosiek, K.Suxho and K.Tamvakis, for valuable discussions.   
This research has been
co-financed by the European Union (European Social Fund - ESF) and Greek national funds
through the Operational Program ''Education and Lifelong Learning" of the National Strategic Reference Framework (NSRF) - Research Funding Program: THALIS-
Investing in the society of knowledge through the European Social Fund.

\end{document}